\documentclass[aps,prd,preprint,draft,showpacs]{revtex4}
\usepackage{graphicx,epsf,amssymb}
\newcommand{\be}{\begin{equation}} 
\newcommand{\ee}{\end{equation}}

\newcommand{\bea}{\begin{eqnarray}} 
\newcommand{\eea}{\end{eqnarray}} 
\newcommand{\Tr}{{\rm Tr}}

\def\V{V}

\newif\ifdraft
\drafttrue
\newif\ifpreprint
\preprinttrue

\def\sect#1{section~{\ref{#1}}}

\def\fig#1{fig.~{\ref{#1}}}

\def\pol{\varepsilon}
\def\Tr{\, {\rm Tr}}

\def\NeqFour{{\cal N}=4}

\def\NA{{\rm NA}}

\def\spa#1.#2{\left\langle#1\,#2\right\rangle}
\def\spb#1.#2{\left[#1\,#2\right]}
\def\sand#1.#2.#3{%
\left\langle\smash{#1}{\vphantom1}^{-}\right|{#2}%
\left|\smash{#3}{\vphantom1}^{-}\right\rangle}
\def\sandp#1.#2.#3{%
\left\langle\smash{#1}{\vphantom1}^{-}\right|{#2}%
\left|\smash{#3}{\vphantom1}^{+}\right\rangle}
\def\sandpp#1.#2.#3{%
\left\langle\smash{#1}{\vphantom1}^{+}\right|{#2}%
\left|\smash{#3}{\vphantom1}^{+}\right\rangle}
\def\sandpm#1.#2.#3{%
\left\langle\smash{#1}{\vphantom1}^{+}\right|{#2}%
\left|\smash{#3}{\vphantom1}^{-}\right\rangle}
\def\sandmp#1.#2.#3{%
\left\langle\smash{#1}{\vphantom1}^{-}\right|{#2}%
\left|\smash{#3}{\vphantom1}^{+}\right\rangle}
\def\sandmm#1.#2.#3{%
\left\langle\smash{#1}{\vphantom1}^{-}\right|{#2}%
\left|\smash{#3}{\vphantom1}^{-}\right\rangle}
\def\spab#1.#2.#3{\sandmm#1.#2.#3}
\def\spba#1.#2.#3{\sandpp#1.#2.#3}
\def\spaa#1.#2.#3.#4{\sandmp#1.{#2#3}.#4}
\def\spbb#1.#2.#3.#4{\sandpm#1.{#2#3}.#4}
\def\spash#1.#2{\spa{\smash{#1}}.{\smash{#2}}}

\newbox\charbox
\newbox\slabox
\def\s#1{{      
        \setbox\charbox=\hbox{$#1$}
        \setbox\slabox=\hbox{$/$}
        \dimen\charbox=\ht\slabox
        \advance\dimen\charbox by -\dp\slabox
        \advance\dimen\charbox by -\ht\charbox
        \advance\dimen\charbox by \dp\charbox
        \divide\dimen\charbox by 2
        \raise-\dimen\charbox\hbox to \wd\charbox{\hss/\hss}
        \llap{$#1$}
}}

\def\eqn#1{eq.~(\ref{#1})}

\def\eqns#1#2{eqs.~(\ref{#1}) and~(\ref{#2})}

\def\qb{{\overline {\kern-0.7pt q\kern -0.7pt}}}

\def\ep{{e^+}}
\def\em{{e^-}}
\def\mup{{\mu^+}}
\def\mum{{\mu^-}}
\def\ib{{\bar\imath}}

\def\prop#1{{\cal P}_{#1}}

\def\sign{{\mathop{\rm sign}\nolimits}}

\def\mod{\mathop{\rm mod}\nolimits}

\def\sandp#1.#2.#3{%
\left\langle\smash{#1}{\vphantom1}^{+}\right|{#2}%
\left|\smash{#3}{\vphantom1}^{+}\right\rangle}
\def\ksl{\s{k}}

\def\proj{{\flat}}

\newbox\ourfigbox
\def\SizedFigureWithCaption#1#2#3{%
\setbox\ourfigbox
  \hbox{\hss\epsfxsize #1 \epsfbox{#2}\hss}
\hbox to \wd\ourfigbox{\vbox{\noindent\copy\ourfigbox\break
\vskip -6mm      \hbox to \wd\ourfigbox{\hss#3\hss}}}
}

\def\spa#1.#2{\left\langle#1\,#2\right\rangle}
\def\spb#1.#2{\left[#1\,#2\right]}
\def\lor#1.#2{\left(#1\,#2\right)}
\def\sand#1.#2.#3{%
\left\langle\smash{#1}{\vphantom1}^{-}\right|{#2}%
\left|\smash{#3}{\vphantom1}^{-}\right\rangle}
\def\sandpp#1.#2.#3{%
\left\langle\smash{#1}{\vphantom1}^{+}\right|{#2}%
\left|\smash{#3}{\vphantom1}^{+}\right\rangle}
\def\sandpm#1.#2.#3{%
\left\langle\smash{#1}{\vphantom1}^{+}\right|{#2}%
\left|\smash{#3}{\vphantom1}^{-}\right\rangle}
\def\sandmp#1.#2.#3{%
\left\langle\smash{#1}{\vphantom1}^{-}\right|{#2}%
\left|\smash{#3}{\vphantom1}^{+}\right\rangle}

\begin{document}
\hfuzz 10 pt


\ifpreprint
\noindent
UCLA/04/TEP/49
\hfill Saclay/SPhT--T04/165
\fi

\title{Twistor-Inspired Construction of Electroweak Vector Boson Currents}%

\author{Zvi Bern}
\affiliation{{} Department of Physics and Astronomy, UCLA\\
\hbox{Los Angeles, CA 90095--1547, USA}
}

\author{Darren Forde} 
\affiliation{Service de Physique Th\'eorique\footnote{Laboratory of the {\it Direction
des Sciences de la Mati\`ere\/} of the {\it Commissariat \`a l'Energie
Atomique\/} of France}, CEA--Saclay\\ 
          F--91191 Gif-sur-Yvette cedex, France
}

\author{David A. Kosower} 
\affiliation{Service de Physique Th\'eorique\footnote{Laboratory of the {\it Direction
des Sciences de la Mati\`ere\/} of the {\it Commissariat \`a l'Energie
Atomique\/} of France}, CEA--Saclay\\ 
          F--91191 Gif-sur-Yvette cedex, France
}

\author{Pierpaolo Mastrolia}
\affiliation{ Department of Physics and Astronomy, UCLA\\
\hbox{Los Angeles, CA 90095--1547, USA}
}

\date{December 13, 2004}

\begin{abstract}
We present an extension of the twistor-motivated MHV vertices and
accompanying rules presented by Cachazo, Svr\v{c}ek and Witten to the
construction of vector-boson currents coupling to an arbitrary
source. In particular, we give rules for constructing off-shell
vector-boson currents with one fermion pair and $n$ gluons of
arbitrary helicity. These currents may be employed directly in the
computation of electroweak amplitudes.  The rules yield expressions in
agreement with previously-obtained results for $Z,W,\gamma^*
\rightarrow \qb q + n$ gluons (analytically up to $n=3$, beyond via
the Berends--Giele recursion relations).  We also confirm that the
contribution to a
seven-point amplitude containing the non-abelian triple vector-boson
coupling obtained using the next-to-MHV currents 
matches the previous result in the literature.
\end{abstract}

\pacs{11.15.Bt, 11.25.Db, 11.25.Tq, 11.55.Bq, 12.38.Bx \hspace{1cm}}

\maketitle



\renewcommand{\thefootnote}{\arabic{footnote}}
\setcounter{footnote}{0}


\section{Introduction}
\label{IntroSection}

Computations of amplitudes in gauge theories play an essential role in
probing beyond the Standard Model of particle physics interactions.
The short-distance experimental environment at hadron colliders forces
us to consider numerous processes with QCD interactions.  Electroweak
gauge bosons have played, and will likely play, an important role
as experimentally distinctive probes of new physics.  Both QCD and
mixed electroweak-QCD processes contribute important backgrounds to
various new-physics signals.  The computation of amplitudes for such
processes is thus an important part of a physics program at modern-day
colliders.  

At tree-level, an elegant --- and computationally-efficient
--- approach was found a decade and a half ago by 
Berends and Giele~\cite{Recurrence}.
Other related approaches have also been 
employed~\cite{OtherNumerical}.  In addition,
various computer-driven approaches such as MADGRAPH~\cite{Madgraph}, are
widely used.

These approaches are not the end of the story, however.  For QCD and
other massless gauge theories, Cachazo, Svr\v{c}ek and Witten
(CSW)~\cite{CSW} have presented a powerful set of new computational
rules for scattering amplitudes.  These rules emerged from analyses
inspired by Witten's~\cite{WittenTopologicalString} proposal of a
weak--weak coupling duality between $\NeqFour$ supersymmetric gauge
theory and the topological open-string $B$ model with a twistor target
space.  This proposal generalizes Nair's earlier
description~\cite{Nair} of the simplest gauge-theory amplitudes.
(Berkovits and Motl~\cite{Berkovits,BerkovitsMotl}, Neitzke and
Vafa~\cite{Vafa}, and Siegel~\cite{Siegel} have given alternative
descriptions of such a possible topological dual to the $\NeqFour$
theory.)  The CSW rules are of interest in their own right for
tree-level computations. Their efficiency can be improved when recast in
a recursive form~\cite{MHVRecursive} (dual in a sense to `connected'
twistor space prescriptions~\cite{Gukov}). They also show great
promise for simplifying loop calculations~\cite{TwistorLoop}.  (A
direct topological-string approach appears to be problematic because
of the appearance of non-unitary states from conformal
supergravity~\cite{BerkovitsWitten}.) Our purpose here is to extend
the CSW construction to include building blocks for mixed
QCD-electroweak amplitudes.

Witten's weak--weak duality is a very promising step towards
understanding why amplitudes in unbroken gauge theories are so much
simpler than expected from the Feynman diagram expansion.  The
best-known example of simplicity in amplitudes is given by the
Parke--Taylor amplitudes~\cite{ParkeTaylor} of massless gauge theories
(including, of course, QCD).  These are an infinite series of
amplitudes, with two negative-helicity gluons and an arbitrary number
of positive-helicity ones (or their parity conjugates). They are
usually called `maximally helicity-violating' (MHV) amplitudes.  They
play an important role in the CSW construction, as the vertices are a
particular off-shell continuation of the MHV amplitudes.  Although
no-one has as yet given a direct derivation of the CSW diagrams from a
Lagrangian there is little doubt they are correct, because they
exhibit the correct kinematic poles and maintain Lorentz invariance,
although it is not manifest~\cite{CSW}.  The rules extend to any
massless gauge theory~\cite{Khoze}, and lead to explicit results for
next-to-MHV amplitudes~\cite{CSW,Khoze,NMHVTree}, and have undergone
non-trivial checks using `googly' amplitudes (the parity conjugates of
the MHV amplitudes) in the context of string and field
theory~\cite{RSV1,OtherGoogly,WittenParity}.  The CSW rules or their
recursive reformulation can be implemented either analytically
or numerically.

The twistor-motivated constructions to date have largely left
unaddressed the question of how to compute processes which contain
both colored and uncolored particles.  Of particular importance to
collider experiments are of course the computation of amplitudes with
both QCD and electroweak particles.  Dixon, Glover, and
Khoze~\cite{LanceHiggs} took the first step in this direction, in
showing how to compute amplitudes containing a Higgs boson coupled to
QCD via a massive top-quark loop (in the infinite-mass limit).  Our
purpose here is to incorporate another important class of processes,
those containing electroweak vector bosons.

We will do so by providing a CSW-like construction of vector currents,
which may in principle be coupled to an {\it arbitrary\/} source.
Typically, one is interested in processes where produced vector bosons
decay into lepton pairs.  For this purpose, one would couple the
currents to lepton pairs.  However, one can also couple them to more
complicated electroweak objects, including, for example, non-abelian
currents decaying in turn into combinations of Higgs bosons and
multiple lepton pairs.  These applications go beyond the CSW
construction, and are a step in the direction of constructing an
off-shell effective action $\Gamma[A]$.

We will focus on the case of coupling a process involving one quark
pair and any number of gluons to one colorless off-shell vector boson.
The constructions should generalize, however, to processes with
multiple quark pairs, and presumably to those with any number of
electroweak gauge bosons.  Indeed, as we shall
discuss, the generalization to symmetrized
electroweak gauge bosons ($Z, \gamma^*$) is straightforward.
It should also be possible to construct
similar currents for {\it colored\/} vector bosons, but we will defer
a discussion of these issues to future papers.  The key idea in the
construction is to introduce a new set of basic vertices coupling to
the off-shell vector boson, analogous to the MHV vertices of CSW.
These basic vertices have either one or no negative-helicity
gluons. The rules for combining them into new currents with additional
negative-helicity legs are then in fact the same as those of CSW.

While we will not provide a first-principles proof of our construction
of the vector boson currents, we shall display considerable evidence
that it is correct.  It exhibits the correct factorization properties
on kinematic poles.  It agrees with previously-computed amplitudes.
In particular, in Appendix C of ref.~\cite{BGKVectorBoson} vector
boson currents for $Z,W,\gamma^* \rightarrow \qb q + n$ gluons up to
$n=3$ are given and we find complete agreement.  We have also
implemented a light-cone gauge version of the Berends-Giele recursion
relations~\cite{Recurrence,LightConeRecursion} and have compared our
new construction numerically with currents computed thereby up to
six gluons of any helicity and eight gluons with selected helicities.

This paper is organized as follows. In \sect{ReviewSection}, we review
color decompositions, especially for the case of vector boson
exchange. In \sect{CSWSection}, we review helicity and MHV
vertices.  In \sect{CurrentsSection}, we generalize MHV vertices to
vertices for currents.  The properties of the amplitudes, including
various consistency checks are discussed in \sect{PropertiesSection}.
Finally, in \sect{ConclusionSection}, we give our conclusions and
outlook.


\section{Review of Color Decompositions with Vector Bosons}
\label{ReviewSection}

It is convenient to write the full momentum-space
amplitudes using color decompositions~\cite{TreeColor,TreeReview}.  
For example, 
the tree-level $n$-gluon amplitude ${\cal A}_n$ has the 
color decomposition,
\begin{equation}
{\cal A}_n(1,2, \ldots, n ) = 
\sum_{\sigma \in S_n/Z_n} \Tr(T^{a_{\sigma(1)}}\cdots T^{a_{\sigma(n)}})\,
A_n(\sigma(1),\ldots,
      \sigma(n)))\,,
\label{TreeColorDecomposition}
\end{equation}
where $S_n/Z_n$ is the group of non-cyclic permutations
on $n$ symbols, and $j$ denotes the $j$-th gluon and
its associated momentum. For now we are suppressing 
the helicity labels.
We use the color normalization $\Tr(T^a T^b) = \delta^{ab}$. 
Similar decompositions hold for cases involving quarks. In general, it is 
more convenient to calculate the partial amplitudes than the
entire amplitude at once.

The cases in which we are interested here involve colorless vector
bosons.  Though it may seem surprising at first sight, single massive
vector boson exchange is easily obtained from pure QCD amplitudes
(which are directly calculable from CSW diagrams).  For example, for
$\ep {\em} \rightarrow \gamma^* \rightarrow q \qb + n$ gluons, where
$\gamma^*$ represents an off-shell photon, the amplitude reduces to
\begin{eqnarray}
\null\hskip -.5 cm 
 {\cal A}_n( 1_\ep, 2_{\em}, 3_q, 4, 5, \ldots, (n-1),  n_\qb)  & = & 
-2 e^2 Q^q 
g^{n-2}\sum_{\sigma \in S_{n-4}} (T^{a_\sigma(4)} T^{a_\sigma(5)} \cdots
    T^{a_\sigma(n-1)})_{i_3}^{\; \ib_n} \nonumber \\
&& \hskip .1 cm \null
 \times  A_n( 1_\ep, 2_{\em}, 3_q, \sigma(4), \sigma(5), \ldots, 
    \sigma(n-1),  n_\qb) \,,
\label{eeZPartons}
\end{eqnarray}
where we use an all outgoing momentum convention.
The particle labels $q, \qb, {\em}, \ep$ signify quarks, anti-quarks,
electrons and positrons, while legs without labels signify gluons.
The off-shell photon, is internal to the amplitude and exchanged
between the lepton pair and the quark pair.

One may then convert the exchanged photon to an electroweak vector
boson very simply by adjusting the coupling and modifying the photon
kinematic pole to be the appropriate one for an unstable massive
particle (see, for example,~\cite{BGKVectorBoson,DKS}). 
In particular, for $ e^+ e^- \rightarrow Z,\gamma^*
\rightarrow \qb q + n g$ one simply modifies the coefficient in
\eqn{eeZPartons} to,
\begin{equation}
Q^q \rightarrow 
 Q^q  - v_{L,R}^e v_{L,R}^q  \, \prop{Z}(s_{12})\,,
\label{PhotonToBoson}
\end{equation}
where $s_{12} = (k_1 + k_2)^2$,
\begin{equation}
\prop{Z}(s_{12}) = {s_{12} \over s_{12} - M_Z^2 + i \,\Gamma_Z \, M_Z}\,,
\end{equation}
and $M_Z$ and $\Gamma_Z$ are the mass and width of the $Z$ boson.
With the replacement~(\ref{PhotonToBoson}), both $Z$ boson and photon
exchange are accounted for in \eqn{eeZPartons}.
The left- and right-handed couplings of fermions to the $Z$ boson are
\begin{eqnarray}
v_L^e & =& { -1 + 2\sin^2 \theta_W \over \sin 2 \theta_W } \,, 
\hskip 2.3 cm 
v_R^e  = { 2 \sin^2 \theta_W \over  \sin 2 \theta_W } \,,  \nonumber \\
v_L^q & =& { \pm 1 - 2 Q^q\sin^2 \theta_W \over  \sin 2 \theta_W } \,,
\hskip 1.9 cm 
v_R^q = -{ 2 Q^q \sin^2 \theta_W \over \sin 2 \theta_W }  \,,
\end{eqnarray}
where $\theta_W$ is the Weinberg angle.  The two signs in
$v_{L,R}^q$ correspond to up $(+)$ and down $(-)$ type quarks.  The
subscripts $L$ and $R$ refer to whether the particle to which the $Z$
couples is left- or right-handed.  That is, $v_R^q$ is to be used for
the configuration where the quark (leg 3) has plus helicity and the
anti-quark (leg $n$) has minus helicity.
Similarly, $v_L^q$ corresponds to the opposite 
configuration.  
For $W$ boson exchange, the propagator correction is
\begin{equation}
\prop{W}(s_{12}) = {s_{12} \over s_{12} - M_W^2 + i \,\Gamma_W \, M_W}\,,
\end{equation}
$M_W$ and $\Gamma_W$ are the mass and width of the $W$ boson,
and the corresponding couplings are,
\begin{equation}
v_R^{f_1\! f_2} = 0\,,  \hskip 2 cm 
v_L^{f_1\! f_2} = {1\over \sqrt{2} \sin\theta_W} U_{f_1\! f_2}
\,.  \
\end{equation}
In the last formula, for quarks $U_{f_1\! f_2}$ is the
Cabibbo-Kobayashi-Maskawa mixing matrix and $f_1$ is an up type 
quark and $f_2$ a down type quark.  For leptons it
is the identity matrix with $f_1$ an up type lepton and $f_2$ a
down type lepton.

More generally, one may convert gluons to photons, purely by group
theoretic rearrangements. (As one example of this, see the fourth
appendix of ref~\cite{QQGGG}.)  However, in general, it is not
possible to then convert the photonic amplitudes to ones involving
electroweak vector bosons since vector bosons have non-abelian self
interactions which photons do not.  Although it is rather pleasing
that the CSW formalism applies directly to single massive vector boson
exchange, cases involving two or more vector bosons are much richer
both from phenomenological and theoretical vantage points. A purpose
of this paper is to construct an appropriate off-shell continuation so
that the CSW methodology can be applied to such cases as well.


\section{Review of CSW Diagrams}
\label{CSWSection}

\def\vo{\vphantom{1}}
The CSW construction~\cite{CSW} builds amplitudes out of vertices
which are off-shell continuations of the Parke--Taylor amplitudes.
These amplitudes, with two negative-helicity
gluons and any number of positive-helicity ones, are the maximally
helicity-violating (MHV) non-vanishing tree-level amplitudes in a
gauge theory.
In the spinor helicity~\cite{SpinorHelicity,XZC,TreeReview}
notation, they are,
\begin{equation}
A_n(1^+,\ldots,m_1^{-},(m_1\!+\!1)\vo^{+},\ldots,m_2^{-},
          (m_2\!+\!1)\vo^{+},\ldots,n^+) = 
  i {\spa{m_1}.{m_2}^4\over \spa1.2\spa2.3\cdots \spa{(n\!-\!1)}.n\spa{n}.1},
\end{equation}
where the two negative-helicity gluons are labeled
$m_{1,2}$. In this equation,
$\spa{i}.{j} = \spa{k_i}.{k_j}$.
We follow the standard spinor normalizations
$\spb{i}.{j} = \sign(k_i^0 k_j^0)\spa{j}.{i}^*$ and
$\spa{i}.{j}\spb{j}.{i} = 2 k_i\cdot k_j$.
With our conventions all particle momenta are taken to be outgoing.

The remaining MHV fermionic amplitudes needed for our discussion
of vector boson currents are,
\begin{eqnarray}
&& A(1_q^+,2^+,3^+, \ldots, i^-, \ldots, (n-2)^+, (n-1)_\qb^-, n^+) =
i {\spa{i}.{1} \spa{i,\,}.{n-1}^3 \over
              \spa1.2 \spa2.3 \spa3.4 \cdots \spa{n}.1}\,, 
\label{MHVtwoquarkphotonD}\\
&& A(1_q^+,2^+,3^+, \ldots, (n-2)^+, (n-1)_\qb^-, n^-) =
i {\spa{n}.{1} \spa{n-1,\,}.n^3 \over
              \spa1.2 \spa2.3 \spa3.4 \cdots \spa{n}.1}\,, 
\label{MHVtwoquarkphotonB} \\
&& A(1_{\qb'}^-,2_{q'}^+, 
    3_q^+, 4^+, \ldots, (n-1)^+, n_\qb^-) 
=  -i {\spa{1}.n^2 \over \spa1.2 \spa3.4 \spa4.5
           \cdots \spa{(n-1)}.{n}} \,. \hskip 1 cm 
\label{MHVfourquarkB}
\end{eqnarray}
The last equation gives the color-ordered amplitude appearing in
\eqn{eeZPartons} after relabeling $q' \rightarrow {\em} $ and
$\qb'\rightarrow \ep$.  These amplitudes may be obtained 
from the purely gluonic MHV ones using supersymmetry identities~\cite{SWI}.

In the CSW construction a particular off-shell continuation of these
amplitudes is an MHV vertex.  The original CSW prescription for the
off-shell continuation of a momentum $k_j$ amounts to replacing
\begin{equation}
\spa{j}.{j'}  \longrightarrow \spb{\eta}.j \spa{j}.{j'}
 \longrightarrow \sandp{\eta}.{\ksl_j}.{j'},
\label{CSWOffShell}
\end{equation}
where $\eta$ is an arbitrary light-like reference vector, in the
Parke-Taylor formula.  The extra factors introduced in this off-shell
continuation cancel when sewing together vertices to obtain an
on-shell amplitude.  As shown by CSW~\cite{CSW}, on-shell amplitudes
are in fact independent of the choice of $\eta$, implying that the
sum over MHV diagrams is Lorentz invariant.

In this paper we use an alternative, but equivalent way of going
off-shell~\cite{NMHVTree,MHVRecursive}.  We instead decompose an
off-shell momentum $K$ into a sum of two massless momenta, where one
is proportional to the auxiliary light-cone reference momentum $\eta$ (with
$\eta^2 = 0$),
\begin{equation}
K = k^\proj + \zeta(K) \eta \, .
\end{equation}
The constraint $(k^\proj)^2 = 0$ yields
\begin{equation}
\zeta(K) =  {K^2 \over 2 \eta\cdot K}.
\end{equation}
If $K$ goes on shell, $\zeta$ vanishes.  Also, if two off-shell vectors
sum to zero, $K_1+K_2=0$, then so do the corresponding $k^\proj$s.
The prescription for continuing MHV amplitudes or vertices off shell is 
to replace,
\begin{equation}
\spa{j}.{j'}\rightarrow \spa{\smash{j^\proj}}.{j'},
\label{OffShellPrescription}
\end{equation}
when $k_j$ is taken off shell.  In the on-shell limit, $\zeta(K)$ vanishes
and $k_j^\proj \rightarrow k_j$.  Although equivalent to the original CSW
prescription, it is a bit more convenient to implement.  In particular,
there are no extra factors associated with going off-shell 
and the MHV vertices carry the same dimensions as amplitudes.

The CSW construction replaces ordinary Feynman diagrams with diagrams
built out of MHV vertices and ordinary propagators.  Each vertex has
exactly two lines carrying negative helicity (which may be on or off
shell), and at least one line carrying positive helicity.  The
propagator takes the simple form $i/K^2$, because the physical state
projector is effectively supplied by the vertices.

We will denote the projected $k^{\proj}$ momentum built out of
$K =k_1 + \cdots k_j$ by $K^\proj$, for example
$\spa{j}.{k^\proj(K,\eta)} = \spa{j}.{K^\proj}$.  
For example, with this notation an all-gluon vertex would
be,
\begin{eqnarray}
&& V(1^+,\ldots,m_1^{-},(m_1\!+\!1)\vo^{+},\ldots, n,
          K\vo^-)\ = i\, 
{\spash{m_1}.{K^\proj}^4 \over \spa1.2\spa2.3\cdots 
      \spash{n}.{K^\proj} 
      \spash{K^\proj}.1} \,.
\label{Vertices}
\end{eqnarray}

For the fermionic vertices, there is a subtlety.  In the standard
phase convention, the on-shell amplitudes have the following 
sort of expression,
\begin{eqnarray}
&& A(1^+,\ldots,f_1^+\ldots,m_1^{-},(m_1\!+\!1)\vo^{+},\ldots, n,
          f_2\vo^-)\ = i\, 
{\spa{m_1}. {f_2}^3 \spa{m_1}.{f_1}
  \over \spa1.2\spa2.3\cdots 
      \spa{n}.{f_2} \spa{f_2}.1} \,.
\label{FermionicAmplitudeA}
\end{eqnarray}
In the CSW approach, the helicity projector in the propagator ---
$\ksl$ in the fermionic case --- is supplied by the vertices.  
In the amplitude, each momentum is directed outwards.  
When we attempt
to link two vertices, we would obtain $|k_1^+\rangle\langle k_2^+|$,
for example, from the numerators.  This is not the correct form,
because $k_1 = -k_2$.  Up to an overall sign, what we want is
$|k_2^+\rangle\langle k_2^+| = |(-k_1)^+\rangle\langle k_2^+|$.
That is, both vertices must have the same
momentum argument in the spinor products.
To correct this, we must flip the sign of
either the positive- or negative-helicity fermion's momentum in the
arguments to the spinor products.  We will choose to flip that
of the negative-helicity fermion, so that the vertex reads,
\begin{eqnarray}
V(1^+,\ldots,f_1^+,\ldots,m_1^{-},(m_1\!+\!1)\vo^{+},\ldots, n,
          f_2\vo^-) & =& -i\, 
{\spa{m_1}. {(-f_2)}^3 \spa{m_1}.{f_1}
  \over \spa1.2\spa2.3\cdots 
      \spa{n}.{(-f_2)} \spa{(-f_2)}.1} \nonumber \hskip 1.3 cm \\
& =& \sign(k_{f_2}^0)\, 
{\spa{m_1}. {f_2}^3 \spa{m_1}.{f_1}
  \over \spa1.2\spa2.3\cdots 
      \spa{n}.{f_2} \spa{f_2}.1} \,,
\label{FermionicVertex}
\end{eqnarray}
where as usual legs are continued off shell by replacing $k_j$ with
$k_j^\proj$, and where $-f_2$ denotes $-k_{f_2}$. 
This issue can be ignored in the gluonic
vertex because there the corresponding factor would involve 
$\sign^2(k^0) = 1$.

The CSW rules using these vertices will yield fermionic amplitude with
a different phase convention than the standard one~\cite{TreeReview}.
To obtain the standard phase conventions, we must multiply the
on-shell amplitude emerging from the CSW rules by a factor of
$i\sign(k^0)$ for the negative-helicity fermion leg.

The CSW rules then instruct us to write down all tree diagrams with
MHV vertices, subject to the constraints that each vertex has exactly
two negative-helicity gluons and at least one positive-helicity gluon
attached, and that each propagator connects legs of opposite helicity.
For amplitudes with two negative-helicity gluons, the vertex with all
legs taken on shell is then the amplitude.  For each additional
negative-helicity gluon, we must add a vertex and a propagator.  The
number of vertices is thus the number of negative-helicity gluons,
less one.

%
\begin{figure}[t]
\centerline{\epsfxsize 4. truein \epsfbox{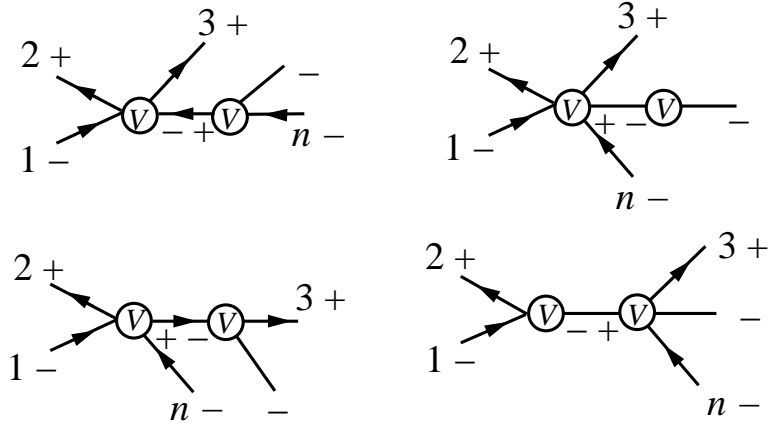}}
\caption{The stripped diagrams for 3 minus helicity amplitudes with
vector boson exchange between two fermion pairs.  Legs 1 and 2
correspond to the leptons and the legs 3 and $n$ to the quarks.
Lines with arrows represent quarks and those without arrows represent
either vector bosons or gluons.}
\label{A4fmmmFigure}
\end{figure}

As a simple example we may use the CSW rules to construct
next-to-MHV (NMHV) partial amplitudes needed for the process $e^+ e^-
\rightarrow \gamma^*, Z, W \rightarrow q \bar q +ng$.
The `stripped diagrams' for this process
are shown in \fig{A4fmmmFigure}.  By stripped diagrams, we mean CSW
diagrams which have been stripped of all the positive helicity gluons.
Dressing the diagrams with the positive helicity gluon
legs between $q$ and $\qb$ in the color ordering leads to
\begin{eqnarray}
&& \hskip -.4 cm 
A(1_{q'}^-, 2_{\qb'}^+,3_q^+, 4^+,5^+, \ldots, (n-1)^-, n_\qb^-)\nonumber\\
&& \hskip 1.7cm 
= \sum_{j = 4}^{n-1}
  V(1_{q'}^-, 2_{\qb'}^+,3_q^+,4^+,5^+, \ldots, (j-1)^+, 
             (-K_{1\ldots (j-1)})_\qb^{-}) \, {i \over K_{1\ldots (j-1)}^2} 
             \nonumber \\
&& \hskip 2.7 cm \null \times
         V((-K_{j\ldots n})_q^{+}, j^+, \ldots, (n-2)^+, 
               (n-1)^-, n_\qb^-) \nonumber \\
&&  \hskip 1.7 cm \null
+  \sum_{j=4}^{n-2}  V(1_{q'}^-, 2_{\qb'}^+,3_q^+,4^+,5^+, \ldots, (j-1)^+, 
                   (-K_{n, 1\ldots (j-1)})^{+}, n_\qb^-) \, 
                             {i \over K_{n,1\ldots (j-1)}^2} \nonumber\\
&& \hskip 2.7 cm \null \times
         V((-K_{j \ldots (n-1)})^{-}, j^+, \ldots, (n-2)^+, 
               (n-1)^-)  \nonumber \\
&&  \hskip 1.7 cm \null
+  V(1_{q'}^-, 2_{\qb'}^+, (-K_{1, 2, n })_q^{+}, n_\qb^-) \, 
                                {i \over K_{n,1,2}^2} \, 
\label{NMHVVectorBosonExchange}  \\
&& \hskip 2.7 cm \null \times
     V(3_q^+, 4^+, \ldots, (n-2)^+,  (n-1)^-,  
                 (-K_{3 \ldots (n-1)})_\qb^{-}) 
                \nonumber \\
&&  \hskip 1.7 cm \null
 +  V(1_{q'}^-, 2_{\qb'}^+, (-K_{1, 2})^{-} ) \, 
                                {i \over K_{1,2}^2} \,  
  V(3_q^+, 4^+, \ldots, (n-2)^+,  (n-1)^-,  n_\qb^-, 
         (-K_{3, \ldots, n})^{+} ) \, ,
\nonumber
\end{eqnarray}
where $K_{i\ldots j} = k_i + k_{i+1} + \cdots + k_j$.  Renaming $\qb',
q' \rightarrow \ep, \em$ gives the partial amplitudes appearing in the
vector boson exchange amplitudes (\ref{eeZPartons}).  As discussed
above, we must multiply by a factor of $i\sign(k_n^0)$ to obtain the
standard phase conventions.

CSW diagrams can be used to obtain any tree-level massless gauge
theory amplitude~\cite{Khoze}.  As the number of negative helicity
legs increases, the complexity of the diagrams increases
rapidly.  As discussed in ref.~\cite{MHVRecursive}, this growth
may be tamed using the recursive rearrangement of diagrams in terms of
non-MHV vertices.

In general, using these amplitudes we may obtain single
vector boson exchange amplitudes by making modifications to the
couplings, color sums and vector boson propagators of the type
discussed in \sect{ReviewSection}.  However, if we wish to apply the
formalism to the phenomenologically interesting cases of vector bosons
decaying into Higgs bosons or multiple fermion pairs including the
non-abelian vector boson self-coupling, an extension of the formalism
is required.  In the next section we describe such an extension.


\section{MHV Vertices for Vector Boson Currents}
\label{CurrentsSection}

In this section we generalize the CSW construction to allow couplings
to arbitrary sources.  We focus on the phenomenologically interesting
case of vector boson currents, though our construction of currents is
applicable more generally.  

An important application of these currents is that they allow us to
couple the electroweak theory to QCD, while taking full advantage of the
CSW formalism on the QCD side.  The currents satisfy a similar color
decomposition as the photon exchange amplitude~(\ref{eeZPartons}),
\begin{eqnarray}
\null\hskip -.5 cm 
 {\cal J}_\mu( 1_q, 2, 3, \ldots, (n-1),  n_\qb; P_\V)  & = & 
g_\V\, g^{n}\sum_{\sigma \in S_{n-2}} (T^{a_\sigma(2)} T^{a_\sigma(3)} \cdots
    T^{a_\sigma(n-1)})_{i_1}^{\; \ib_n} \nonumber \\
&& \hskip .1 cm \null
 \times  J_\mu( 1_q, \sigma(2), \sigma(3), \ldots, 
    \sigma(n-1),  n_\qb ; P_\V) \,,
\label{CurrentColor}
\end{eqnarray}
where $g_\V$ is the appropriate coupling for a vector boson $\V=\gamma^*,Z,W$
and $P_\V$ is the momentum carried by the vector boson.
Hence we need consider only the partial currents $J_\mu$ in much 
the same way that we need only consider color-ordered partial amplitudes.

We start by defining two currents that will serve
as new basic vertices for obtaining general vector boson currents:
\begin{enumerate}

\item A vector-boson current with $n$ gluon emissions, all of positive
helicity
\begin{eqnarray}
\null \hskip -.5 cm 
J^\mu(1_q^-,2^+, \ldots, (n-1)^+, n_\qb^+; P_\V) 
  &=& -{i \over \sqrt{2}} {\sandmp{(-1)}.{\gamma^\mu \s P_\V}.{(-1)}
 \over \spa{(-1)}.2 \spa2.3 \ldots \spa{(n-1)}.{n} } \nonumber\\
 &=&  c_+ \, \pol^{(+)\mu} (P_\V^\proj,\eta)
     +c_- \, \pol^{(-)\mu} (P_\V^\proj,\eta) \nonumber \\
&& \hskip .5 cm \null 
  + c_L \, \biggl( P_\V^\mu - {P_\V^2\over \eta\cdot P_\V} \eta^\mu\biggl)\,, 
\hskip 3 cm 
\label{BasicFermionicCurrent}
\end{eqnarray}
where $P_\V = -K_{1\ldots n}$ by momentum conservation, where `$-1$'
as a spinor argument denotes $-k_1$, and where
\begin{eqnarray}
c_+ &=& -V^{\rm MHV}(1_\qb^-,\ldots,n_q^+;P_\V^-)\,,\nonumber\\
c_- &=& V^{\rm MHV}(1_\qb^-,\ldots,n_q^+;P_\V^-) \,
             {\spa1.{\eta}^2 P_\V^2\over
              \spash{\eta}.{P_\V^\proj}^2\spash1.{P_\V^\proj}^2}\,,
\label{BasicFermionCoefficients}\\
c_L &=& V^{\rm MHV}(1_\qb^-,\ldots,n_q^+;P_\V^-) \,
{\sqrt2\spa1.{\eta}
 \over\spash{\eta}.{P_\V^\proj}\spash1.{P_\V^\proj}} \,.\nonumber
\end{eqnarray}
The vertex $V^{\rm MHV}$ is simply a CSW vertex for one photon, one
quark pair, and $n-2$ gluons, obtained by fermionic phase adjustments
from the amplitude in \eqn{MHVtwoquarkphotonB}.

If the helicity assignments of the fermions are reversed
we instead have
\begin{equation}
\null \hskip -.5 cm 
J^\mu(1_q^+,2^+, \ldots, (n-1)^+, n_\qb^-; P_\V) 
  = -{i \over \sqrt{2}} {\sandmp{(-n)}.{\gamma^\mu \s P_\V}.{(-n)}
  \over \spa1.2 \spa2.3 \ldots \spa{(n-1)}.{(-n)} }\,,
\label{BasicFermionicCurrentB}
\end{equation}
as well as a similar decomposition in polarizations.  As with the
basic CSW vertices, when any colored leg $j$ is taken off shell, the $k_j$
argument to all spinor products or spinor strings must be replaced by
$k_j^\proj$.  (The vector momentum $P_V$ is already off shell, and should
{\it not\/} be replaced by $P_V^\proj$ if the latter is not already present
in the above formul\ae{}.)

\item A purely bosonic basic current emitting a single vector
state,
\begin{equation}
J^\mu((-P_\V)^-;P_\V) 
  = {i \over \sqrt{2}} {\sandpp{\eta}.{\gamma^\mu}.{P_\V^\proj}
                         \over \spb{P_\V^\proj}.{\eta}} P_\V^2
   = i \, \pol^{(-)\mu}(P_\V^\proj, \eta) P_\V^2 \,.
\label{BasicGluonicCurrent}
\end{equation}
\end{enumerate}
The first of these is the vector-boson current for positive
helicity gluons~\cite{BGKVectorBoson}. The second is just a negative
helicity polarization vector with reference momentum taken to be the
CSW reference momentum.

The polarizations in the above equations are defined using
the spinor helicity method and are given by~\cite{XZC} 
\begin{equation}
\pol_\mu^{(+)}(k, r) = 
 {1\over \sqrt{2}} {\sandmm{r}.{\gamma^\mu}.{k} \over \spa{r}.{k}}\,,
\hskip 2 cm 
\pol_\mu^{(-)}(k, r) = 
 {1\over \sqrt{2}} {\sandpp{r}.{\gamma^\mu}.{k} \over \spb{k}.{r}}\,,
\end{equation}
where $r$ is a null reference momentum.
When evaluating expressions involving these polarizations, useful
identities are  
\begin{equation}
\sandpp{a}.{\gamma^\mu}.{b} =  \sandmm{b}.{\gamma^\mu}.{a} \,,
\end{equation}
and the Fierz identity,
\begin{equation}
\sandmm{a}.{\gamma^\mu}.{b} \sandpp{c}.{\gamma_\mu}.{d} =  2
 \spa{a}.{d} \spb{c}.{b} \,.
\label{FierzIdentity}
\end{equation}

We take the currents (\ref{BasicFermionicCurrent}) and
(\ref{BasicGluonicCurrent}) to act as vertices, using the same CSW
prescriptions (\ref{CSWOffShell}) or (\ref{OffShellPrescription}) as
used for defining vertices from MHV amplitudes.  The denominator of
the first current (\ref{BasicFermionicCurrent}) contains only angle
spinor products. That is, it depends only on on spinors $\lambda$ and
not on conjugate spinors $\tilde \lambda$, and hence is holomorphic in
the spinor variables.  Accordingly, this current will have
derivative-of-delta-function support in twistor
space~\cite{WittenTopologicalString}.  The denominator of 
the second current (\ref{BasicGluonicCurrent}) contains a bracket spinor
product.  It is thus analogous to the non-NHV amplitude vertices
defined in ref.~\cite{MHVRecursive}. To obtain general currents with an
arbitrary number of negative helicity legs we follow the same CSW
rules for amplitudes, except that now we have two new vertices.

\begin{figure}[t]
\centerline{\epsfxsize 6 truein \epsfbox{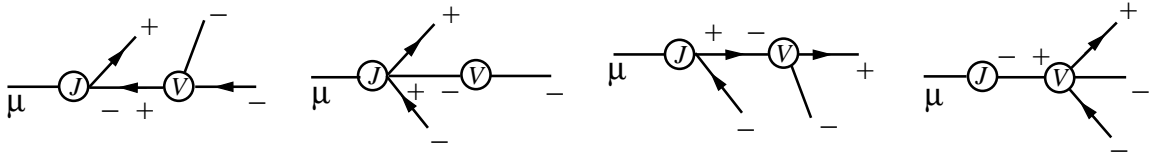}}
\caption[a]{\small 
The NMHV vector boson current in terms of diagrams where
positive helicity gluon lines have been stripped.
}
\label{JfmmFigure}
\end{figure}

To illustrate the construction of a current with more negative
helicities, consider the NMHV vector boson current,
$J_{\mu}(1_q^+,2^+,\ldots,(n-2)^+,(n-1)^-,n^{-}_{\qb};P_\V)$
where the negative helicity legs are nearest neighbors in the
color ordering.  The CSW diagrams for this current may be organized
using the four diagrams shown in \fig{JfmmFigure},
where the positive helicity gluon legs have all been stripped away.
Inserting back the positive helicity gluon legs, leads to the following
expression for this NMHV vector boson current,
\begin{eqnarray}
&&
\hskip -1 cm 
J_{\mu}(1_q^+,2^+,\ldots,(n-2)^+,(n-1)^-,n^{-}_{\qb}; P_\V)
\nonumber\\
&& \hskip 1 cm 
=\sum_{j=2}^{n-1}J_{\mu}(1_q^+,2^+,\ldots,(j-1)^+,
  (K_{j\ldots n})^{-}_{\qb}; P_\V)
  \frac{i}{K_{j \ldots n}^2}
\nonumber \\
&& \hskip 2 cm \null \times
V( (-K_{j\ldots n})_{q}^{+}, 
    j^+,\ldots,(n-2)^+,(n-1)^-,n^-_{\qb})
\nonumber \\
&&\hskip 1.2 cm \null 
+\sum_{j=2}^{n-2}J_{\mu}(1_q^+,2^+,\ldots,(j-1)^+,
               (K_{j\ldots (n-1)})^{+},n^-_{\qb}; P_\V )
\frac{i}{K_{j\ldots (n-1)}^2}
\nonumber \\
&& \hskip 2.5 cm \times \null 
V((-K_{j \ldots (n-1)})^{ -} ,j^+,\ldots,(n-2)^+,(n-1)^-)
\nonumber \\
&& \hskip 1.2 cm \null
+J_{\mu}((K_{1 \ldots (n-1)})^{ +}_{q},n^-_{\qb}; P_\V)
\frac{i}{K_{1 \ldots (n-1)}^2}
\label{NMHVCurrent} \\
&&  \hskip 2.5 cm \times
V(1_q^+,2^+,\ldots,(n-2)^+,(n-1)^-, (-K_{1 \ldots (n-1)})_{\qb}^{ -})
\nonumber \\
&& \hskip 1.2 cm \null 
+ J_{\mu}((K_{1 \ldots n})^- ; P_\V )
\frac{i}{K_{1 \ldots n}^2}
V(1_q^+,2^+,\ldots,(n-2)^+,(n-1)^-,n_{\qb}^-, (-K_{1\ldots n})^{+})\,,
\nonumber \end{eqnarray}
where the momentum of the vector boson is $P_\V = - K_{1\ldots n}$.
The explicit values of the current vertices are obtained from
\eqns{BasicFermionicCurrentB}{BasicGluonicCurrent} by relabeling the
arguments.  Other NMHV helicity configurations are only a bit more
complicated.

In ref.~\cite{MHVRecursive}, Bena and two of the authors introduced
a recursive reformulation of the CSW rules, useful when increasing
the number of negative helicity legs.  It introduces
a higher-degree composite vertex, $V_n$, carrying $c$ negative helicities
defined in terms of two simpler off-shell vertices and a propagator via,
\begin{eqnarray}
&& V_n(1^+,\ldots,m_1^-,(m_1\!+\!1)\vo^+,\ldots,
m_2^-,(m_2\!+\!1)\vo^+,\ldots,
m_c^-,(m_c\!+\!1)\vo^+,\ldots,n^+) = \nonumber \\
&&\hskip 7mm
{1\over (c-2)}\sum_{j_1=1}^n\mathop{{\sum}'}_{j_2=j_1}^{j_1-3}
{i\over K_{j_1\ldots j_2}^2} V_{j_2-j_1+2 \mod n}
 (j_1,\ldots,j_2,(-K_{j_1\ldots j_2})\vo^{-}) 
\label{VertexRecursion}\\ 
&& \hskip 4mm \hphantom{ {1\over (c-2)}\sum_{j_1=1}^n\sum_{j_2=j_1+1}^{j_1-3}
                           {i\over K_{j_1\ldots j_2}^2 } }\hskip 6mm\times
V_{j_1-j_2\mod n}(j_2+1,\ldots,j_1-1,(-K_{(j_2+1)\ldots(j_1-1)})\vo^+)\,,
\nonumber
\end{eqnarray}
where each term is included only if there is at least one 
negative-helicity gluon in the cyclic range $[j_1,j_2]$ and at least two
in the range $[j_2+1,j_1-1]$. The prime on the sum indicates
the omission of any term with vanishing denominator.  The subscript on
$V$ indicates the number of colored legs.
All indices are
to be understood $\mod n$, and all sums in a cyclic sense,
for example,
\begin{equation}
\sum_{j=n-4}^3 \equiv \sum_{j={(n-4)\cdots n,1\cdots3}}\qquad
{\rm\ and\ } \qquad \sum_{j=2}^{-2} \equiv \sum_{j=2}^{n-2}.
\label{CyclicSumDef}
\end{equation}

The sums here have been modified from those presented in 
ref.~\cite{MHVRecursive} to include two-point vertices, which
however vanish for purely colored legs.
The two vertices in each term are then
simpler in the sense that each has lower degree,
that is fewer 
negative-helicity legs (including the off-shell ones) than the parent
vertex.  This means that this equation provides a recurrence relation
for evaluating any tree-level non-MHV vertex in massless gauge theory,
and therefore for evaluating any tree-level current or amplitude.
Note that the sum implicitly runs over different degrees
for the two vertices on the right-hand side, because the number of
negative-helicity external legs in $[j_1,j_2]$ can vary.

Indeed, the {\it same\/} recurrence relation also applies to the
vector-boson currents considered in the present paper, so long as we
interpret one of the indices as being the free vector index, and we
define its `helicity' to be negative for bookkeeping purposes inside
the sums. The vector index must be adjacent to both quark legs in the
cyclic ordering.  (Now there is a non-vanishing two-point current to
consider, which was the motivation for modifying the limits of the
sums.  In accordance with~\eqn{BasicGluonicCurrent}, the non-vector
leg on this vertex also has two negative-helicity legs.  Two-point
vertices without a vector index are taken to vanish.)

Alternatively, we could write out separate terms corresponding to
higher-degree vertices analogous to those shown in
\fig{JfmmFigure}. This would give us an expression that is a
generalization of \eqn{NMHVCurrent}. The first three terms will be
double sums; the currents and vertices will be higher-degree objects;
and there will be an overall combinatoric factor as in \eqn{VertexRecursion}.

Extending this construction to multi-vector currents is straightforward
when they are commuting ($Z,\gamma^*$), because we can then symmetrize
on the vector legs.  In this case, the corresponding basic vertices
include those in \eqns{BasicFermionicCurrent}{BasicGluonicCurrent},
as well as additional vertices constructed from \eqn{BasicFermionicCurrent}
by replacing the one-photon vertex $V^{\rm MHV}$ of
\eqn{BasicFermionCoefficients} with a multi-photon vertex, 
$V^{\rm MHV}(1_\qb^-,\ldots,n_q^+;P_{\V1}^-,P_{\V2}^+,\ldots)$, where one
photon is taken to have negative helicity, and the remainder, positive
helicity.  In \eqn{VertexRecursion}, each vector-current index should be
counted as a negative helicity for book-keeping purposes.


\section{Properties of the Currents}
\label{PropertiesSection}

The currents obtained using the vertices
defined in the previous section and using CSW diagrams ought to
satisfy a number of
non-trivial properties, such as current conservation.  In addition,
because as colorless vector-boson currents they are gauge invariant
(think of them as off-shell photon currents),
we may suspect and that they will
satisfy Berends-Giele recursion relations.  They indeed do.
Because the vector boson
currents are gauge invariant they should also match the previously-computed
vector-boson currents with up to three gluons, as given in appendix~C of
ref.~\cite{BGKVectorBoson}.  By sewing together two currents with a
vector boson propagator, we can obtain vector-boson
exchange amplitudes.  Moreover, since the vector-boson leg is fully
off shell it may be coupled to electroweak Feynman diagrams, enhancing
the applicability of CSW diagrams to include processes mixing QCD and
electroweak theory.

The simplest requirement that any current should satisfy is
conservation.  With our construction, the conservation of currents
with arbitrary gluon helicities follows directly from current
conservation of the basic currents (\ref{BasicFermionicCurrent}) and
(\ref{BasicGluonicCurrent}); their conservation is unaffected by the
CSW off-shell continuations (\ref{CSWOffShell}) or
(\ref{OffShellPrescription}).

The current must also yield an amplitude when the off-shell leg 
is contracted with a photon polarization vector, and the on-shell
limit is taken.  For example, contract
$\pol^{(+)}_\mu(k_{n+1}, r)$ into the NMHV current (\ref{NMHVCurrent}),
where $r$ is an arbitrary null reference momentum and take
the limit $P_\V = -K_{1\ldots n} \rightarrow  k_{n+1}$, with $k_{n+1}^2 = 0$.
 In this case, all but the last term in
the current (\ref{NMHVCurrent}) give vanishing contributions because, 
\begin{eqnarray}
\pol_{\mu}^{(+)}(k_{n+1}, r)  \, 
J^{\mu}(1^+_q,2^+,\ldots,(j-1)^+, (-K_{1 \ldots j-1})^{-}_{\qb};
           P_\V) = 0\,,
\label{VanishingDotProduct}
\end{eqnarray}
using \eqn{BasicFermionicCurrentB} and the Fierz 
identity (\ref{FierzIdentity}).
In the  last term in \eqn{NMHVCurrent} we obtain
\begin{equation}
 \pol^{(+)}_\mu(k_{n+1}, r) \, J^{\mu}( (-P_V)^- ; P_\V)
\rightarrow  -i K_{1\ldots n}^2 \,,
\end{equation}
using \eqns{BasicGluonicCurrent}{FierzIdentity}, which  cancels
the factor of  $i/ K_{1\ldots n}^2$ in the last term of \eqn{NMHVCurrent}.
The remaining factor is just the desired amplitude, so we have,
\begin{eqnarray}
\hskip -.3 cm 
&& \pol_\mu^{(+)}(k_{n+1}, r) \,
 J^{\mu}(1_q^+,2^+,\ldots,(n-2)^+,(n-1)^-,n^{-}_{\qb}; P_\V)
 \nonumber \\
&& \hskip 2.8 cm 
\rightarrow A(1_q^+,2^+,\ldots,(n-2)^+,(n-1)^-, n^{-}_{\qb}, (n+1)^+)\,,
\label{PolPlusCurrent}
\end{eqnarray}
which is the required property.

Similarly, contracting $\pol^{(-)}_{\mu}(k_{n+1}, r)$ with the current
(\ref{NMHVCurrent}) gives
\begin{eqnarray}
&&
\hskip -.8 cm 
\pol_\mu^{(-)}(k_{n+1}, r)  
J^{\mu}(1_q^+,2^+,\ldots,(n-2)^+,(n-1)^-,n^{-}_{\qb};P_\V)
\nonumber\\
&& \hskip 1cm 
\rightarrow \sum_{j=2}^{n-1}
V(1_q^+,2^+,\ldots,(j-1)^+,(K_{j\ldots n})^-_{\qb}, (n+1)^- )
\frac{i}{K_{j \ldots n}^2}
\nonumber \\
&& \hskip 2.5 cm \null \times
V( (-K_{j\ldots n})_{q}^{+},
        j^+,\ldots,(n-2)^+,(n-1)^-,n^-_{\qb})
\nonumber \\
&&\hskip 1.2 cm  \null 
+\sum_{j=2}^{n-2} V(1_q^+,2^+,\ldots,(j-1)^+,
  (K_{j\ldots (n-1)})^{+}, n^-_{\qb}, (n+1)^-)
\frac{i}{K_{j\ldots n-1}^2}
\nonumber \\
&& \hskip 2.5 cm \null \times
V( (-K_{j\ldots (n-1)})^{ -} ,j^+,\ldots,(n-2)^+,(n-1)^-)
\nonumber \\
&& \hskip 1.2 cm \null  
+ V((-K_{n,(n+1)})^{+}_{q},n^-_{\qb}, (n+1)^-)
\frac{i}{K_{1 \ldots (n-1)}^2}
\nonumber \\
&&  \hskip 2.5 cm \null \times
V(1_q^+,2^+,\ldots,(n-2)^+,(n-1)^-, (-K_{1\ldots (n-1)})_{\qb}^{-}) \,.
\label{NMHV2QAmplitude}
\end{eqnarray}
This may be recognized as the CSW expression for an NMHV amplitude,
confirming that
\begin{eqnarray}
&& \pol_\mu^{(-)}(k_{n+1}, r) 
J^{\mu}(1_q^+,2^+,\ldots,(n-2)^+,(n-1)^-,n^{-}_{\qb};P_\V) \nonumber\\
&& \hskip 6 cm 
\rightarrow
 A(1_q^+,2^+,\ldots,(n-2)^+,(n-1)^-,n^{-}_{\qb}, (n+1)^-)\,.  \hskip 1 cm 
\end{eqnarray}

Following similar algebraic steps, it is then straightforward to
demonstrate that for all possible helicity choices for gluons $2, \ldots
(n-1)$, and independent of the number of MHV vertices in each diagram,
\begin{eqnarray}
&& \pol_\mu^{(\pm )}(k_{n+1}, r) 
J^{\mu}(1_q^+,2,\ldots,(n-2),(n-1),n^{-}_{\qb}; P_\V) \nonumber\\
&& \hskip 6 cm 
= A(1_q^+,2,\ldots,(n-2),(n-1),n^{-}_{\qb}, (n+1)^\pm)\,. \hskip 1 cm 
\end{eqnarray}

Another property that currents should satisfy is that a product of
two currents contracted with a vector boson propagator should give
amplitudes with colorless vector boson exchange. 
For example, from the MHV currents (\ref{BasicFermionicCurrent}) 
and~(\ref{BasicFermionicCurrentB}) 
we may obtain the 
MHV partial amplitudes needed for vector boson exchange between a lepton 
and quark pair, given in \eqn{eeZPartons}, 
\begin{eqnarray}
&&  \hskip -1.5 cm 
A(1_{\ep}^-, 2_{\em}^+,3_q^+, 4^+, \ldots, (n-1)^+, n_\qb^-) \nonumber \\
&& \hskip 2 cm =  J_\mu(1_{\ep}^-, 2_{\em}^+; -P_{1,2})\,
{i \over P^2_{1,2}} \, 
J^\mu (3_q^+, 4^+, \ldots, (n-1)^+, n_\qb^-; P_{1,2}) \,,
\label{BasicFermionicFactorization}
\end{eqnarray}
where $P_{1,2} = k_1 + k_2$.  This relation between the MHV amplitudes
and currents continues to hold when the on-shell legs are taken off
shell using the CSW prescriptions~(\ref{CSWOffShell})
or~(\ref{OffShellPrescription}).  Using
\eqns{BasicFermionicFactorization}{NMHVCurrent} it is then straightforward
to verify that amplitudes with larger numbers of negative-helicity gluons
satisfy similar relations. For example, 
\begin{equation}
J_\mu(1_{\qb'}^-, 2_{q'}^+; -P_{1,2})
{i \over P^2_{1,2}} \, 
J^\mu (3_q^+, 4^+, \ldots, (n-1)^-, n_\qb^-; P_{1,2})\,,
\label{NMHVFermionicFactorization}
\end{equation}
reproduces the CSW expression for the NMHV single vector boson exchange
amplitude (\ref{NMHVVectorBosonExchange}), as required.

\begin{figure}[t]
\centerline{\epsfxsize 5 truein \epsfbox{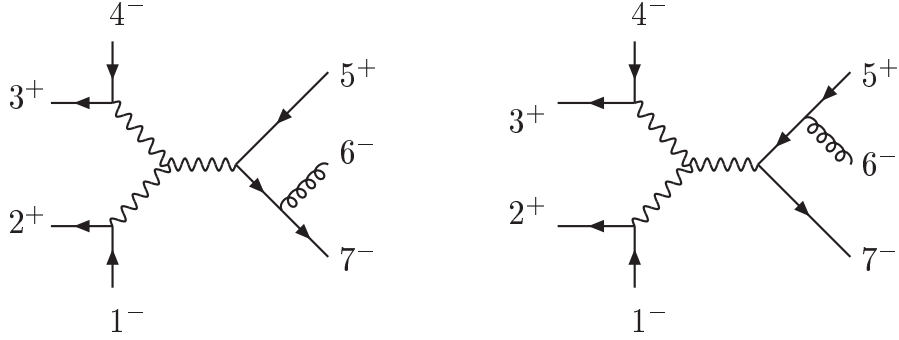}}
\caption[a]{\small 
The two Feynman diagrams contributing the seven point amplitude with a
triple vector boson self coupling. Wavy lines represent gauge bosons
such as the $W$ or $Z$ and curly lines gluons.}
\label{ATripleBosonFigure}
\end{figure}

A key property of the vector boson currents is that they may be
coupled to off-shell electroweak gauge bosons.  For example, consider
the amplitude containing the non-abelian triple vector boson self
coupling ({\it i.e.} the $ZW^+W^-$ coupling), $A_7^{\NA} (1^-_{\em},
2^+_\ep, 3^+_\mum, 4^-_\mup, 5_q^+, 6^-, 7_\qb^-)$, depicted in terms
of Feynman diagrams in \fig{ATripleBosonFigure}.  Up to relabelings
and a parity reflection, it corresponds to the amplitude in eq.~(2.23)
of ref.~\cite{DKS}.  As explained in that reference, the sum over
these two diagrams is gauge invariant because its electroweak coupling
constant differs from diagrams without a triple vector boson
self-interaction and hence do not mix with them.

\begin{figure}[t]
\centerline{\epsfxsize 1.8 truein \epsfbox{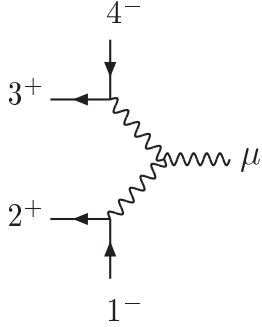}}
\caption[a]{\small 
The Feynman diagram for the electroweak vector boson current with
a non-abelian triple vector boson self interaction.  We may couple
this to the QCD vector boson current determined using CSW diagrams.}
\label{JTripleBosonFigure}
\end{figure}

In our construction this amplitude factorizes into a product of two
currents with the connecting propagator.  One is the QCD vector boson
current described by the NMHV current $J_{\mu}(5_q^+,6^-,7^-_{\qb};
P_\V)$ obtained by relabeling \eqn{NMHVCurrent}.  The
electroweak current containing the non-abelian triple vector boson
interaction may easily be evaluated by working out the Feynman diagrams
shown in \fig{JTripleBosonFigure} in Feynman-'t~Hooft or any other
gauge.  We have verified that combining the two currents reproduces
eq.~(2.23) of ref.~\cite{DKS} (after accounting for a parity reflection
and relabeling).  Of course, this example is sufficiently
simple so that MHV diagrams are unnecessary.  However,
with the CSW approach it is straightforward to incorporate larger
numbers of quarks and gluons in the vector boson QCD current.

The complete process also includes contributions in which both
vector bosons couple directly to the fermion line, given in eqs.~(2.8)
and~(2.22)
of ref.~\cite{DKS}.  We have verified that a symmetrized double-vector
current constructed as described in the previous section reproduces
the results of ref.~\cite{DKS} after symmetrization over the two leptonic 
currents $(3,4)\leftrightarrow(6,5)$.

We have implemented a light-cone version of the Berends--Giele 
recurrence relations, for the purposes of confirming our extension of
CSW diagrams to currents.  These are obtained by reshuffling terms
from the standard ones (similar to the spirit of 
ref.~\cite{LightConeRecursion}). We have used these to check
currents computed using the recursive reformulation of the CSW
rules, as described in \eqn{VertexRecursion}.  We performed this comparison
numerically, for all helicities up to $n=8$ legs (in addition
to the leg carrying the vector-boson index).  We also checked
selected helicities through $n=10$.

\section{Conclusions and Outlook}
\label{ConclusionSection}

The twistor-inspired computational approach presented by Cachazo,
Svr\v{c}ek, and Witten is a novel way of computing tree amplitudes in
massless gauge theories, including of course QCD.  It relies on a
basic set of vertices which are localized in a twistor-space sense.  These
vertices are particular off-shell continuations of the well-known 
Parke--Taylor amplitudes.  The CSW approach does not address the computation
of amplitudes containing both colored and non-colored particles.  In
the context of the Standard Model, it does not, for example address
the question of computing mixed QCD-electroweak processes.

In this paper, we have taken a step in this direction.  
We have shown how to incorporate
an additional vector leg coupling to an arbitrary source into the
CSW approach.  As we have discussed, multiple symmetrized vector legs
($Z,\gamma^*$) follow with a straightforward modification.
The propagators and diagrammatic rules are essentially
unchanged; it suffices to add new basic vertices with the additional
vector leg(s).  (Unlike the CSW construction, these vertices do include
a two-point vertex.)  In our explicit examples, we have focused on
coupling one off-shell electroweak vector boson to a single quark pair and any
number of gluons.  The structure of the CSW construction implies that
adding additional quark pairs should be straightforward.  
We expect that a similar approach can be used to construct
multi-$W$s currents, and that 
ideas presented in this paper also generalize to currents for
{\it colored\/} particles.  These topics will be elaborated 
on elsewhere.

To date, no-one has given a complete first-principles proof of the CSW
construction.  Nonetheless, there is very strong evidence in its
favor.  As pointed out in the original paper~\cite{CSW}, a CSW
computation automatically has the correct collinear and
multiparticle-factorization limits.  In addition, extensive numerical
checks against Berends--Giele recurrence relations (which were derived
from first principles) through relatively high multiplicity leave
little room for doubt.

The factorization arguments also apply to the currents presented in
this paper.  We have implemented a light-cone version of the
Berends--Giele recurrence relations~\cite{BGKGluon}, and have shown
complete agreement between currents computed using them, and currents
computed using a recursive~\cite{MHVRecursive} reformulation of the
CSW rules.  We performed these comparisons numerically, for up to ten legs
(in addition to the leg carrying the vector-boson index).  Photon
amplitudes can be derived from the currents (by contracting with a
polarization vector, and taking the on-shell limit), and we have shown
that these expressions reproduce the corresponding amplitudes as
obtained directly from the CSW rules, after adjusting the color
factors to reflect the abelian nature of the photon.

The currents we have constructed can be used directly in the
computation of processes producing electroweak vector bosons.
We are optimistic that further phenomenologically useful
extensions of the CSW approach will be possible.

\section*{Acknowledgments}

We would like to thank Emil Bjerrum-Bohr, Lance Dixon and Dave Dunbar 
for helpful
discussions.  P.M. would also like to thank the University of Zurich
for its kind hospitality while this paper was being completed.
This research was supported by the US Department of
Energy under contract DE--FG03--91ER40662.


\end{document}